\documentclass[12pt]{article}
\newcommand{\e}[1]{\epsilon_{#1}}
\def\baselinestretch{1.0}
\catcode`\@=11
\def\@normalsize{\@setsize\normalsize{15pt}\xiipt\@xiipt
\abovedisplayskip 14pt plus3pt minus3pt%
\belowdisplayskip \abovedisplayskip
\abovedisplayshortskip  \z@ plus3pt%
\belowdisplayshortskip  7pt plus3.5pt minus0pt}
\def\small{\@setsize\small{13.6pt}\xipt\@xipt
\abovedisplayskip 13pt plus3pt minus3pt%
\belowdisplayskip \abovedisplayskip
\abovedisplayshortskip  \z@ plus3pt%
\belowdisplayshortskip  7pt plus3.5pt minus0pt
\def\@listi{\parsep 4.5pt plus 2pt minus 1pt
            \itemsep \parsep
            \topsep 9pt plus 3pt minus 3pt}}
\def\underline#1{\relax\ifmmode\@@underline#1\else
        $\@@underline{\hbox{#1}}$\relax\fi}
\@twosidetrue \relax \catcode`@=12
\catcode`\@=11

\def\section{\@startsection{section}{1}{\z@}{3.5ex plus 1ex minus
   .2ex}{2.3ex plus .2ex}{\large\bf}}

\def\ps@headings{\def\@oddfoot{}\def\@evenfoot{}
\def\@oddhead{\hbox{}\hfill
        \makebox[.5\textwidth]{\raggedright\ignorespaces --\thepage{}--
        \hfill }}
\def\@evenhead{\@oddhead}
\def\subsectionmark##1{\markboth{##1}{}}
} \ps@headings \catcode`\@=12 \relax

\def\figcap{\section*{Figure Captions\markboth
        {FIGURECAPTIONS}{FIGURECAPTIONS}}\list
        {Fig. \arabic{enumi}:\hfill}{\settowidth\labelwidth{Fig. 999:}
        \leftmargin\labelwidth
        \advance\leftmargin\labelsep\usecounter{enumi}}}
 \relax
\def\tablecap{\section*{Table Captions\markboth
        {TABLECAPTIONS}{TABLECAPTIONS}}\list
        {Table \arabic{enumi}:\hfill}{\settowidth\labelwidth{Table 999:}
        \leftmargin\labelwidth
        \advance\leftmargin\labelsep\usecounter{enumi}}}
 \relax
\def\reflist{\section*{References\markboth
        {REFLIST}{REFLIST}}\list
        {[\arabic{enumi}]\hfill}{\settowidth\labelwidth{[999]}
        \leftmargin\labelwidth
        \advance\leftmargin\labelsep\usecounter{enumi}}}
 \relax
\catcode`\@=11
\def\marginnote#1{}
\newcount\hour
\newcount\minute
\newtoks\amorpm
\hour=\time\divide\hour by60 \minute=\time{\multiply\hour by60
\global\advance\minute by- \hour}
\edef\standardtime{{\ifnum\hour<12 \global\amorpm={am}%
    \else\global\amorpm={pm}\advance\hour by-12 \fi
    \ifnum\hour=0 \hour=12 \fi
    \number\hour:\ifnum\minute<100\fi\number\minute\the\amorpm}}
\edef\militarytime{\number\hour:\ifnum\minute<100\fi\number\minute}
\def\draftlabel#1{{\@bsphack\if@filesw {\let\thepage\relax
  \xdef\@gtempa{\write\@auxout{\string
    \newlabel{#1}{{\@currentlabel}{\thepage}}}}}\@gtempa
    \if@nobreak \ifvmode\nobreak\fi\fi\fi\@esphack}
     \gdef\@eqnlabel{#1}}
\def\@eqnlabel{}
\def\@vacuum{}
\def\draftmarginnote#1{\marginpar{\raggedright\scriptsize\tt#1}}
\def\draft{\oddsidemargin -.5truein
        \def\@oddfoot{\sl preliminary draft \hfil
        \rm\thepage\hfil\sl\today\quad\militarytime}
        \let\@evenfoot\@oddfoot \overfullrule 3pt
        \let\label=\draftlabel
        \let\marginnote=\draftmarginnote
\def\@eqnnum{(\theequation)\rlap{\kern\marginparsep\tt\@eqnlabel}%
\global\let\@eqnlabel\@vacuum}  }
\def\preprint{\twocolumn\sloppy\flushbottom\parindent 1em
        \leftmargini 2em\leftmarginv .5em\leftmarginvi .5em
        \oddsidemargin -.5in    \evensidemargin -.5in
        \columnsep 15mm \footheight 0pt
        \textwidth 250mmin      \topmargin  -.4in
        \headheight 12pt \topskip .4in
        \textheight 175mm
        \footskip 0pt
\def\@oddhead{\thepage\hfil\addtocounter{page}{1}\thepage}
        \let\@evenhead\@oddhead \def\@oddfoot{} \def\@evenfoot{}
}
\def\titlepage{\@restonecolfalse\if@twocolumn\@restonecoltrue\onecolumn
     \else \newpage \fi \thispagestyle{empty}\c@page\z@
        \def\thefootnote{\fnsymbol{footnote}} }
\def\endtitlepage{\if@restonecol\twocolumn \else  \fi
        \def\thefootnote{\arabic{footnote}}
        \setcounter{footnote}{0}}  %\c@footnote\z@ }
\catcode`@=12 \relax
\def\ps@headings{\def\@oddfoot{}\def\@evenfoot{}
\def\@oddhead{\hbox{}\hfill
        \makebox[.5\textwidth]{\raggedright\ignorespaces --\thepage{}--
        \hfill }}
\def\@evenhead{\@oddhead}
\def\subsectionmark##1{\markboth{##1}{}}
} \ps@headings \relax
\newcommand{\newc}{\newcommand}

\newc{\ra}{\rightarrow}
\newc{\lra}{\leftrightarrow}
\newcommand{\ba}{\begin{eqnarray}}
\newcommand{\ea}{\end{eqnarray}}
\def\tila{\tilde{\alpha}}

\usepackage{epsfig}

\begin{document}
\def\firstpage#1#2#3#4#5#6{
%%%%%%%%%%%
\begin{titlepage}
\nopagebreak
\title{\begin{flushright}
\vspace*{-0.8in}
{ \normalsize  hep-ph/yymmddd\\
%CERN-TH /2002\\
%  IOA-20/2002 \\
%December 2002 \\
}
\end{flushright}
\vfill {#3}}
\author{\large #4 \\[1.0cm] #5}
\maketitle \vskip -7mm \nopagebreak
\begin{abstract}
{\noindent #6}
\end{abstract}
\vfill
\begin{flushleft}
\rule{16.1cm}{0.2mm}\\[-3mm]
%%%%%%%%%%%
\end{flushleft}
\thispagestyle{empty}
\end{titlepage}}

\def\simlt{\stackrel{<}{{}_\sim}}
\def\simgt{\stackrel{>}{{}_\sim}}
\date{}
\firstpage{3118}{IC/95/34} {\large\bf Gauge coupling and fermion
mass relations in low string scale brane models} {D.V.~Gioutsos,
G.K. Leontaris and J. Rizos} {\normalsize\sl Theoretical Physics
Division, Ioannina University,
GR-45110 Ioannina, Greece\\[2.5mm]}
{We analyze the gauge coupling evolution  in  brane inspired models
with $U(3)\times U(2)\times U(1)^N$ symmetry at the string scale.
We restrict to the case  of brane
configurations with two and three abelian factors ($N=2,3$) and
 where {\it only one} Higgs doublet is coupled to down
quarks and leptons and only one to the up quarks. We  find
that the  correct hypercharge assignment of the Standard Model particles
is reproduced  for six viable models distinguished by different brane configurations.
 We investigate the third generation fermion mass relations  and find that the
correct low energy $m_b/m_{\tau}$ ratio can be obtained for $b-\tau$
Yukawa coupling equality at a string scale as low as $M_S\sim 10^3\,$ TeV.}

\def\baselinestretch{1.3}
\section{Introduction}

Low scale unification of gauge and gravitational interactions
~\cite{Antoniadis:1990ew,Arkani-Hamed:1998rs,Antoniadis:1998ig}
appears to be a promising framework for solving the hierarchy
problem. In this context, the weakness of the gravitational force in
long distances is attributed to the existence of extra dimensions at
the Fermi scale. A realization of this scenario can occur in type I
string theory~\cite{Lykken:1996fj} where gauge interactions are
mediated by open strings with their ends attached on some D-brane
stack, while gravity is mediated by closed strings that propagate in
the whole 10--dimensional space.

In the context of Type I string theory using appropriate collections
of parallel\footnote{For reviews see~\cite{Polchinski:1996na},
\cite{Angelantonj:2002ct}.} or
intersecting~\cite{Berkooz:1996km,Balasubramanian:1996uc} D-branes,
there has been considerable work in trying to derive the Standard
Model theory or its Grand Unified extensions
\cite{Antoniadis:2002en}-\cite{Blumenhagen:2005mu}.
Some of these low energy models revealed rather interesting
features: (i) The correct value of the weak mixing angle
can be reproduced for a string scale of the order of a few TeV (ii) baryon and
lepton numbers are conserved due to the existence of  exact global
symmetries which are remnants of additional anomalous $U(1)$ factors
broken by the Green-Schwarz mechanism (iii) supersymmetry is not
necessary  to solve the hierarchy problem.

However, its rivals,  supersymmetric Grand Unified theories and
their heterotic string realizations exhibit additional interesting
features as: (i) full gauge coupling unification, that occurs to a
scale of the order $10^{16}$ GeV or a little higher when additional
matter thresholds are introduced (see eg. \cite{Kaplunovsky:1987rp})
(ii)  fermion mass relations \cite{Chanowitz:1977ye,Greene:1986jb}
and in particular bottom-tau unification, i.e., the equality of the
corresponding Yukawa couplings at the unification scale, which
reproduces the correct mass relation $m_b/m_{\tau}$ at low energies.

 In Type I string scenarios the volume of the internal space
enters in the relation between gauge and string couplings
\cite{Witten:1996mz} and in general  predictability is lost.
However, there exist classes of models where (due to some internal
volume relation in intersecting brane models or superposition of the
associated parallel brane sets) at least the two of the brane
couplings are either related  or equal at the string scale $M_S$
(e.g., ``petite
unification"\cite{Leontaris:2001hh,Kiritsis:2003mc}). For models
with three  gauge group factors one such relation is enough to
associate the low energy data, i.e., the Weinberg angle and the
strong coupling, with the string scale. Given the variety of the
type I vacua one could follow a bottom-up approach and consider
ratios of brane couplings as free parameters. In this context, some
restrictions on the parameter space can be obtained assuming certain
fermion mass relations at the string scale. We will concentrate here
on restrictions implied by the heaviest generation fermion mass
relations and in particular the bottom - tau Yukawa couplings at the
string scale.

Following a bottom-up approach, in this letter  we examine  the
possible brane configurations that can accommodate the Standard
Model and the associated hypercharge embeddings and we analyze the
consequences of (partial) gauge coupling unification in conjunction
with bottom-tau coupling equality. We shall restrict
 to non-super\-symmetric configurations, (for some recent results on
 supersymmetric models see \cite{Blumenhagen:2003jy}), however,
 we will consider models
with two Higgs doublets so that the bottom and top quark masses
will be related to different vacuum expectation values while  the
tau lepton and the bottom quark will  receive masses from the same
Higgs doublet. We find that in a class of models that can be
realized in the context of type I string theory with large extra
dimensions, the experimentally low energy masses can be reproduced
assuming equality of bottom-tau Yukawa couplings and a string
scale as low as $M_S \sim 10^3$ TeV.

In the next section we briefly describe the general set up of
brane models and identify candidate brane configurations that
allow  bottom-tau coupling equality. All possible hypercharge
embeddings in the presence of additional $U(1)$ factors are
classified.  Section 3 deals with the calculational details and
renormalization analysis of gauge and Yukawa couplings, while in
section 4  the results for $b-\tau$ couplings are presented. Our
conclusions are drawn in section 5.

\section{Hypercharge embedding in generic Standard model like brane configurations}

In this work, we consider models which arise in the context of
various D-brane configurations \cite{Antoniadis:2002en,
Antoniadis:2002qm}. A single D-brane carries a $U(1)$ gauge symmetry
which is the result of the reduction of the ten-dimensional
Yang-Mills theory. Therefore, a stack of $n$ parallel, almost
coincident D-branes gives rise to a $U(n)$ gauge theory where the
gauge bosons correspond to open strings having both their ends
attached to some of the branes of the various stacks.

The minimal number of brane sets  required to provide the Standard
Model structure is three: a 3-brane ``color" stack with gauge
symmetry ${U(3)}_C\sim{SU(3)}_C\times{U(1)}$ gives rise to strong
interactions, a 2-brane ``weak" stack  gives rise to
${U(2)}_L\sim{SU(2)}_L\times{U(1)}$ gauge symmetry that includes the
weak interactions and an abelian $U(1)$ brane for hypercharge.
However, accommodation of  all SM particles as open strings between
different brane sets requires at least one $U(1)$ brane to be added
to the above
configuration~\cite{Antoniadis:2002en,Antoniadis:2002cr}.
 Additional abelian branes may be present too. In
more complicated scenarios the weak or color stacks can be repeated
leading to an effective ``higher level embedding" of the Standard
Model.
The full gauge group will be of the form
\ba
G={U(m)}_C^p\times{U(n)}_L^q\times{U(1)}^N \label{ggg}
 \ea
 with
$m\ge 3$ and $n\ge 2$ and $p,q\ge1$.  Since $U(n)\sim
{SU(n)}\times{U(1)}$ and so on, we infer that brane constructions
automatically give rise to models with $SU(n)$ gauge group structure
and several $U(1)$ factors.

A generic feature of this type of string vacua is that several
abelian gauge factors are anomalous. Note that this is in contrast
to the heterotic case where at most one $U(1)$ is anomalous,
however, anomalies are cancelled by a generalized Green-Schwarz
mechanism. At least one $U(1)$ combination  remains anomaly free.
This is the hypercharge that can be in general written as \ba
Y=\sum_{i=1}^p k_3^{(i)} Q_3^i + \sum_{j=1}^q k_2^{(j)} Q_2^j+
\sum_{\ell=1}^N k'_\ell \, Q'_\ell, \label{ydef} \ea where $Q_3^i$
are the $U(1)$ generators of the color factor $i$, $Q_2^j$ are the
${U(1)}$ generators of the weak factor $j$ and
$Q'_\ell\,(\ell=1,\dots,N)$, are the generators of the remaining
Abelian factors.

The  simplest case which leads directly to the SM theory is the
choice $p=q=1$. Constructions  of this type  have already been
proposed in the papers of reference~\cite{Antoniadis:2002en}. An immediate
consequence of (\ref{ggg}) and (\ref{ydef})  is that the hypercharge
coupling ($g_Y$) at the string/brane scale $(M_S)$ is related to the
brane couplings ($g_m, g_n, g_i'$) as\footnote{If some Cartan
generators of $SU(m)$ also contribute to the hypercharge, the
formula becomes model dependent, see e.g. \cite{Leontaris:2001hh}.}
\ba
\frac{1}{g_Y^2}=\frac{2 m k_3^2}{g_m^2}+\frac{2 n
k_2^2}{g_n^2}+2\sum_{i=1}^N \frac{{k'_i}^2}{ {g'_i}^2}\label{gydef}
\ea
where we have used the traditional normalization ${\rm
Tr}\,T^a\,T^b=\delta^{ab}/2, a,b=1,\dots,n^2$ for the $U(n)$
generators and assumed that the vector representation (${\bf n}$)
has abelian charge $+1$ and thus the $U(1)$  coupling becomes
${g_n}/{\sqrt{2 n}}$ where $g_n$ the $SU(n)$ coupling.

Choosing further $m=3, n=2$ in (\ref{gydef}) we obtain directly the
non-abelian structure of the SM with several $U(1)$ factors,
therefore the hypercharge gauge coupling condition reads
\ba
\frac{1}{g_Y^2}=\frac{6 k_3^2}{g_3^2}+\frac{4
k_2^2}{g_2^2}+2\sum_{i=1}^N \frac{{k'_i}^2}{{g'_i}^2}\label{gY}
\ea
For a given hypercharge embedding the $k_i'$'s are known and
equation (\ref{gY}) relates the weak angle with the
gauge coupling ratios at the string scale
\ba
\sin^2\theta_W(M_S)&=&\frac{1}{1+k_Y}\label{sin2w}\\
k_Y&\equiv&\frac{\alpha_2}{\alpha_Y}
   \;=\;{6 k_3^2}\,\frac{\alpha_2}{\alpha_3}+4 k_2^2+2\sum_{i=1}^N
{k_i'}^2\,\frac{\alpha_2}{\alpha_i'}\label{kY} \ea where $\alpha_i
\equiv g_i^2/(4\pi)$.

Given a relation between the $\alpha_i'$ and $\alpha_2$ (or
$\alpha_3$),  equations (\ref{sin2w},\ref{kY}) in conjunction with
the $\alpha_3$ evolution equation, determine the string scale $M_S$.
In the remaining of this section,  we will  derive  all possible
sets of $k_i$'s compatible with brane configurations which embed the
SM particles and imply an economical Higgs spectrum.

\begin{figure}[!ht]
\centering
($N=2$) \includegraphics[width=0.28\textwidth]{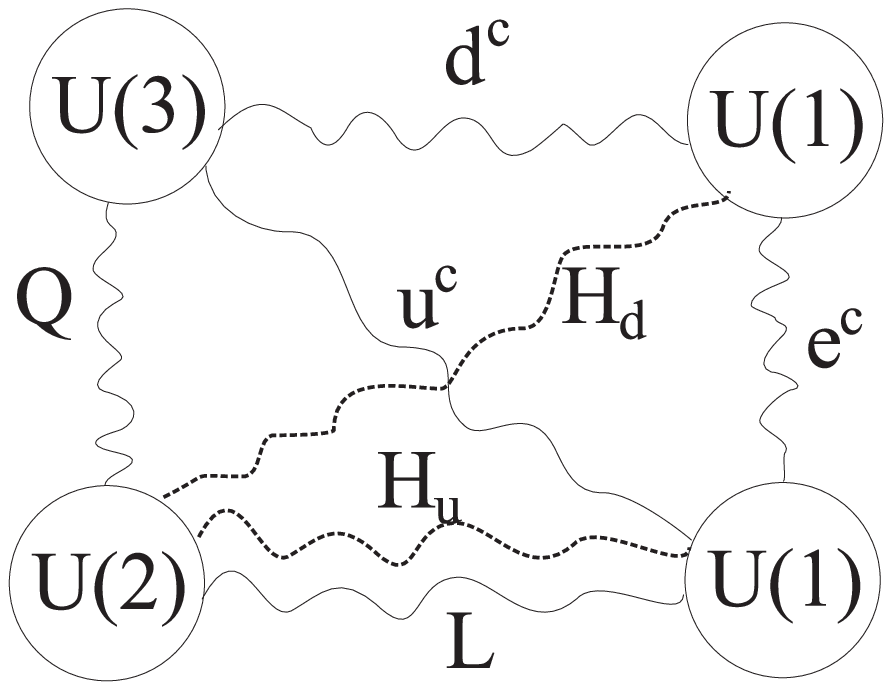}\ \ \ \ \ \ \ \ \ \ \ \ \ \ \ ($N=3$)
\includegraphics[width=0.25\textwidth]{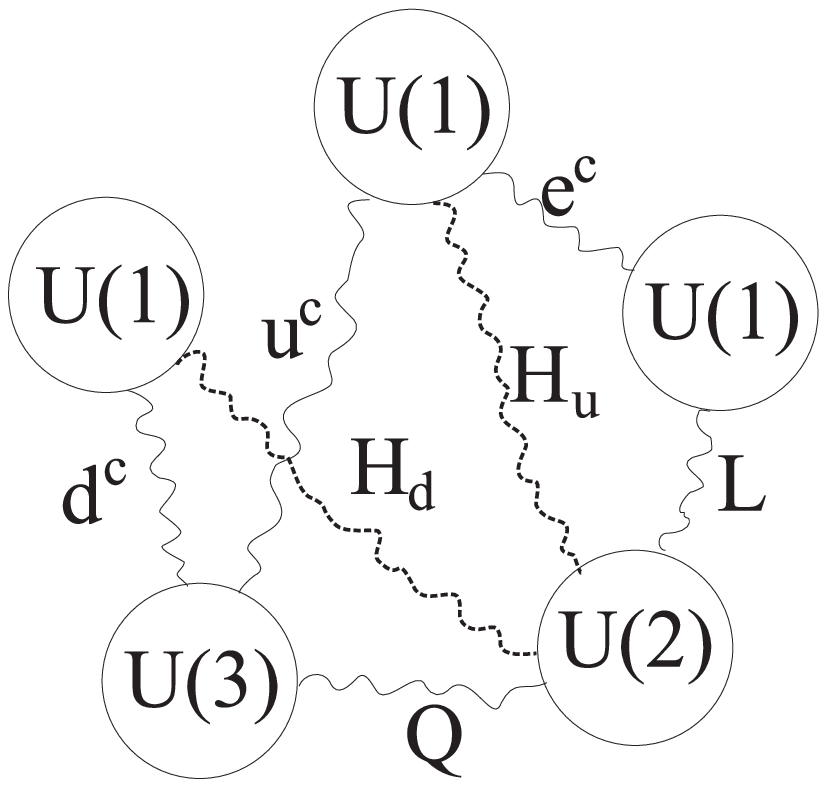}
\caption{\label{fconf} Possible $N=2,3$ brane configurations ($N$
is the number of the $U(1)$-branes) that can accommodate the SM
spectrum with down quarks and leptons acquiring masses from the
same Higgs doublet. The first consists of four brane sets and has
gauge symmetry $U(3)\times U(2)\times {U(1)}^2$ and the second
consists of five brane sets and has gauge group $U(3)\times
U(2)\times {U(1)}^3$.}
\end{figure}

In brane models each SM particle  corresponds to an open string
stretched between two branes. In our charge conventions, the
possible quantum numbers of such a string ending to the $U(m)$ and
$U(n)$ brane sets are $\left({\bf m};+1,{\bf n};+1\right),\left({\bf
\overline{m}};-1,{\bf n};+1\right)$, $\left({\bf m};+1,{\bf
\overline{n}};-1\right)$, $\left({\bf \overline{m}};-1,{\bf
\overline{n}};-1\right)$, that is, bi-fundamentals of the associated
unitary groups. Higher antisymmetric or symmetric representations
could also be obtained by considering strings with both ends on the
same brane set, however,  we will restrict here to the
bi-fundamental case. In order to ensure $\lambda_b-\lambda_{\tau}$
unification at $M_S$, we assume models where  down quarks and
leptons acquire their masses from a common Higgs.  Only two
configurations ($N=2,3$) share the above properties and are
presented pictorially in Fig. \ref{fconf}.

The possible hypercharge embeddings  for each configuration can be
obtained by solving the hypercharge assignment conditions for SM
particles. Generically, the SM particle abelian charges under
$U_3(1) \times U_2(1) \times {U(1)'}_1 \times {U(1)'}_2$ are of the
form $Q(+1,\epsilon_1,0,0)$, $d^c(-1,0,\epsilon_2,0)$,
$u^c(-1,0,0,\epsilon_3)$, $L(0,\epsilon_4,0,\epsilon_5)$,
$e^c(0,0,\epsilon_6,\epsilon_7)$, where $\e{i}=\pm1, i=1,\dots,7$.
Then, the SM hypercharge assignment conditions read
 \ba
k_3+k_2\, \e1 &=& \hphantom{+}\frac{1}{6}\nonumber\\
-k_3+k_1'\, \e2 &=& \hphantom{+}\frac{1}{3}\nonumber\\
-k_3+k\, \e3 &=& -\frac{2}{3}\label{6e}\\
k_2\, \e4+k_2'\, \e5 &=& -\frac{1}{2}\nonumber\\
k_1'\,\e6 +k_2'\, \e7 &=& \hphantom{+}1\nonumber
\ea
for $Q$, $d^c$, $u^c$, $L$ and $e^c$ respectively.
 Here we have used a compact notation where $k=k_2'$
for the first configuration and $k=k_3'$ for the second one.
%Solving for $\left|k_i\right|,
%\left|k_i'\right|$ we obtain the first three solutions of Table
%\ref{ytab1} in the four-brane stack scenario and the next four in
%the five-brane stack one.
\footnote{In some of our solutions there exist additional
unbroken $U(1)$'s factors. We will assume in the sequel
that these $U(1)$'s will be broken by vacuum expectation values of
additional SM singlet Higgs fields or at the string level by
six-dimensional anomalies~\cite{Antoniadis:2002cr}.}

As seen by  (\ref{ydef}) and (\ref{gydef}), only the absolute values
of the hypercharge embedding coefficients $k_i,\,k_i'$ enter the
coupling relation at $M_S$. Solving (\ref{6e}), for the SM particle
charges in configuration ($N=2$) we obtain three possible solutions.
These correspond to the (absolute) values for the coefficients
presented in cases (a),(b) and (c) of Table \ref{ytab1}.
Configuration $N=3$ leads to four additional cases, namely (d),
(e),(f) and (g) of the same Table. If in a particular solution a
coefficient $k_i$ (or $k_i'$) turns out to be zero, the associated
abelian factor does not participate to the hypercharge.
\begin{table}[!ht]
\centering
\begin{tabular}{|c|l|c|c|c|c|c|}
\hline
$N$&&$|k_3|$&$|k_2|$&$|k_1'|$&$|k_2'|$&$|k_3'|$\\
\hline
&(a)&$\frac{1}{6}$&$0$&$\frac{1}{2}$&$\frac{1}{2}$&-\\
$2$&(b)&$\frac{2}{3}$&$\frac{1}{2}$&$1$&$0$&-\\
&(c)&$\frac{1}{3}$&$\frac{1}{2}$&$0$&$1$&-\\
\hline
&(d)&$\frac{1}{6}$&$0$&$\frac{1}{2}$&$\frac{1}{2}$&$\frac{1}{2}$\\
&(e)&$\frac{1}{3}$&$\frac{1}{2}$&$0$&$1$&$1$\\
$3$&(f)&$\frac{5}{6}$&$1$&$\frac{1}{2}$&$\frac{1}{2}$&$\frac{3}{2}$\\
&(g)&$\frac{2}{3}$&$\frac{1}{2}$&$1$&$0$&0\\
\hline
\end{tabular}
\caption{\label{ytab1}Absolute values of the possible hypercharge
embedding coefficient sets ($k_3, k_2$ and $k_i'$)  for the brane configurations with $N=2$ and $N=3$
of Figure \ref{fconf}. }
\end{table}

\section{Gauge coupling running and the String scale}

As already mentioned, in the low energy Type I string scenarios the gauge
couplings
%$g_3, g_2$
 do not unify at the string scale $M_S$.
%Similarly the couplings $g'_i$ of the abelian gauge factors are not
%necessarily the same at $M_S$.
\footnote{One may think that the predictability of these
constructions is lost, however, bearing in mind that the origin of
each gauge factor is due to a different stack of branes, the
situation of unequal couplings at the effective string scale,
although not predictive, looks completely natural.} However, it has
been observed that in some cases low energy data are compatible with
a  partially unified model where some of the gauge couplings are
equal (``petite
unification"~\cite{Leontaris:2001hh,Kiritsis:2003mc}). At the string
level this scenario corresponds to superposing the associated
parallel brane stacks. Moreover certain coupling relations arise in
classes of intersecting brane models. Given the fact that there may
exist various gauge coupling relations at the string scale $M_S$
(although, only one for a minimal model), low energy electroweak
data can be used to determine $M_S$ through the renormalization
group equations (RGEs). Following this bottom-up approach, in this
section we determine the range of the string scale for all the above
models by taking into account the experimental values of $\alpha_3,
\alpha_{em}$ and $\sin^2\theta_W$  at  $M_Z$ \cite{Eidelman}
 \ba
\alpha_3 = 0.118\pm 0.003,~~~\alpha^{-1}_{e}=127.906,~~~
\sin^2\theta_W=0.23120\nonumber \ea For the scales above $M_Z$ we
consider the standard model spectrum with two Higgs doublets. The
one loop RGEs  for the gauge couplings ($\tilde{\alpha}\equiv
\alpha/(4\pi)$) take the form \ba \frac{d \tilde{\alpha}_i}{dt} =
b_i \tilde{\alpha}_i^2\,,~~~~ i=Y,2,3\label{rge} \ea where $(b_Y,
b_2, b_3)= (7, -3, -7)$ and $t=2\ln\mu$ ($\mu$ is the
renormalization point).

\begin{table}[!ht]
\centering
\begin{tabular}{|c|c|c|c||c|c|c|}
\hline
&\multicolumn{6}{|c|}{Model}\\
\hline%
\parbox{1cm}{\small coupling\\[-6pt]relation}&(a)&(b)&(c)&(d)&(e)&(g)\\
\hline%
 $\alpha_i'=\alpha_2$&$\frac{\xi}{6}+1$&$\frac{8\xi}{3}+3$ &
$\frac{2\xi}{3}+3$ & $\frac{\xi}{6}+\frac{3}{2}$ & $\frac{2\xi}{3}+5$ & $\frac{8\xi}{3}+3$ \\
\hline%
$\alpha_i'=\alpha_3$&$\frac{7\xi}{6}$ & $\frac{1\bf{4\xi}}{3}+1$ & $\frac{8\xi}{3}+1$ &
$\frac{2\xi}{3}+1$ & $\frac{14\xi}{3}+1$ & $\frac{14\xi}{3}+1$ \\
\hline%
 (see text)&$\frac{2\xi}{3} +\frac{1}{2}$ & - & - & $\frac{5\xi}{3}$ & $\frac{8\xi}3+3$ & - \\
\hline
(see text)&- & - & - & $\frac{7\xi}{6}+\frac 12$ & - & - \\
\hline
\hline
$M_S $(GeV)&$6.71\times 10^{17}$&$5.78\times 10^3$&$1.99\times 10^6$&$1.65\times 10^{14}$
&$5.78\times 10^3$&$5.78\times 10^3$\\
\hline
\end{tabular}
\caption{\label{ytab2} The possible values of $k_Y$ as a function
of $\xi=\frac{\alpha_2}{\alpha_3}$ for various orientations of
$U(1)$'s for the models of Table \ref{ytab1}. The first row presents
the $k_Y$ values when all $U(1)$ branes are aligned with the
$SU(2)$ brane, i.e., when $\alpha_i'=\alpha_2$, while the second one
corresponds to $\alpha_i'=\alpha_3$. The next two rows show the $k_Y$
values for other possible orientations (see text for details). Last row
shows the minimum value of the string scale $M_S$ obtained for the models $a-g$.}
\end{table}

First, we concentrate on simple relations of the gauge couplings,
i.e., those relations implied from  models arising only in the
context on non-intersecting branes.  In these cases, certain
constraints on the initial values of the gauge couplings have to be
taken into account, leading to a discrete number of admissible cases
which we are going to discuss. Thus,  in the case of two $D5$
branes, $U(3)$ and $U(2)$ are confined in different bulk directions.
In the parallel brane scenario the orientation of a number of the
extra $U(1)$'s may coincide with the $U(3)$-stack direction while
the remaining abelian branes are parallel to the $U(2)$ stack. This
implies that the corresponding $U(1)$ gauge couplings have the same
initial values either with the $\alpha_3$ or with the $\alpha_2$
gauge couplings. If we define $\xi=\frac{\alpha_2}{\alpha_3}$ the
ratio of the two non-abelian gauge couplings at the string scale,
for any distinct case,  $k_Y$ takes the form $k_Y= \lambda \,\xi
+\nu$, where $\lambda,\nu$ are calculable coefficients which depend
on the specific orientation of the $U(1)$ branes. For example, in
model (a) we can have the following possibilities:
$\alpha_1'=\alpha_2'=\alpha_2$, $\alpha_1'=\alpha_2'=\alpha_3$ and
$\alpha_1'=\alpha_2, \alpha_2'=\alpha_3$ leading to
$k_Y=\frac{\xi}6+1$, $\frac{7\xi}6$ and $\frac{2\xi}3+\frac 12$
correspondingly. All cases for the models (a-g)  are presented in
Table~\ref{ytab2} and are classified with regard to the hypercharge
coefficient $k_Y$. (The analysis shows that all cases of Model (f)
lead to unacceptably small string scales, so these are not
presented). Allowing ${\alpha_3}$ to take values different from
${\alpha_2}$, we find that models (a,b,c,d,e,g) of Table \ref{ytab1}
predict a string scale in a wide range, from a few TeV up to the
Planck mass. The highest value  is of the order $M_S\sim 7\times
10^{17}$ GeV and corresponds to equal couplings
$\frac{\alpha_2}{\alpha_3}\equiv \xi =1$ at $M_S$. On the other
hand, lower unification values of the order of a few  TeV assume a
gauge coupling ratio $\frac{\alpha_3}{\alpha_2}\approx 2$. In this
case the idea of complete gauge coupling unification could be still
valid,  considering that the SM gauge group arises from the breaking
of a gauge symmetry whose non-abelian part is $U(3)\times U(2)^2$,
i.e., for the case $p=1,\,q=2$ of (\ref{ggg}) where the factor of 2
in the gauge coupling ratio is related to the diagonal breaking
$U(2)\times U(2)\ra U(2)$. The distinct cases with the predictions
for the unification scale and other quantities we are interested in,
are presented in the columns (2-4) of Table~\ref{ytab3a}. The lowest
possible unification for the three models
 (b),(e),(g) corresponds to  $k_Y=\frac{14\,\xi}3+1$, and is $M_S\sim
5.81\times 10^{3}$ GeV, for a weak to strong gauge coupling ratio
$\xi\sim 0.42$ at $M_S$. Case (c) predicts an intermediate value
$M_S =2 \times 10^6$ GeV while model (d) gives $M_S\sim 10^{14}$
GeV. Finally, model (a) for $\xi\sim 1$ predicts a unification scale
as high as $M_S\sim 6.7\times 10^{17}$ GeV which is of the order of
the heterotic string scale. Interestingly, in this latter case, all
gauge couplings are equal at $M_S$, $\alpha_3=\alpha_2=\alpha_i'$,
while, as can be seen from Table 2, $k_Y$ takes a common value for
all three cases, $k_Y=7/6$.

\begin{figure}[!ht]
\centering%
\epsfxsize=12cm \epsfbox{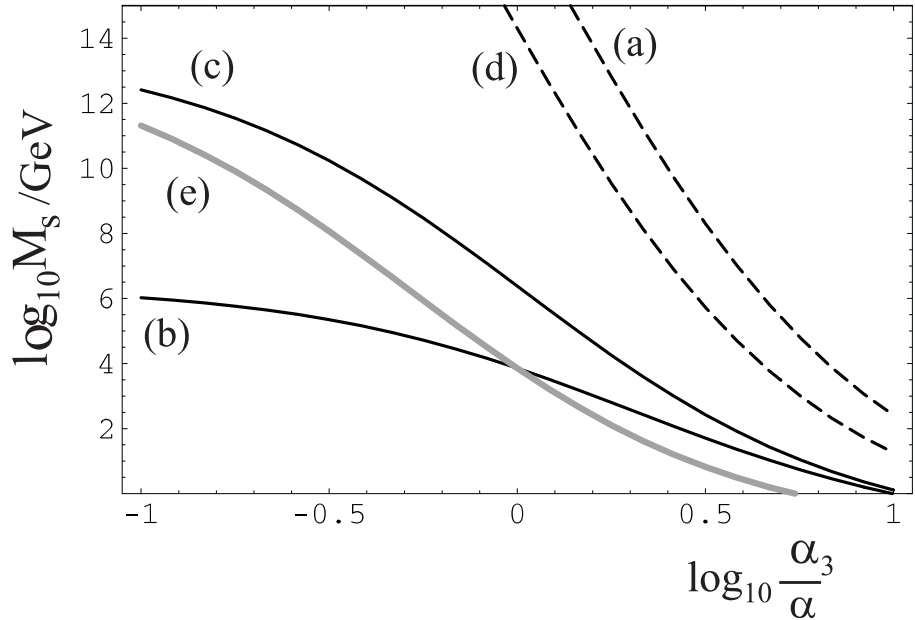} \caption{\label{cmodels}The
string scale as a function of the coupling ratio
$\frac{\alpha_3}{\alpha}$, ($\alpha$ is a common value for the
$U(1)$ couplings $\alpha_i'$) for the different hypercharge
embeddings of Table 1, in the general case of intersecting branes.
The results for model (g) are identical with those of model (b).}
\end{figure}

In the general intersecting case, the $U(1)$ branes are neither
aligned to the $SU(3)$, nor to the $SU(2)$ stacks, thus the
corresponding gauge couplings can take arbitrary values.  Without
loss of generality, we will assume here for simplicity that all
these couplings are equal
$\alpha_1'=\alpha_2'=\dots=\alpha_N'=\alpha$. In Figure
\ref{cmodels} we plot the string scale ($M_S$) as a function of the
logarithm of the ratio $\alpha_3/\alpha$ for the candidate models
(a), (b), (c), (d), (e) and (g).  The results for models (b), (c),
(e) and (g) which is identical with model (b), are represented in
the figure with continuous lines. These are compatible with low
scale unification particularly when $\alpha_3\ge \alpha$. For
$\alpha_i'=\alpha_3$, (which corresponds to the zero of the
logarithm at the $x$-axis), we obtain again the results of the
parallel brane scenario, shown in Table 2. At this point, we further
observe a crossing of the (e)-curve with the curve for models
(b),(g). It is exactly this point ($\alpha_i'=\alpha_3$) that these
three models predict the same value for the lowest string scale.
When $\alpha_3\ge \alpha_i'$, model (e) predicts the lowest $M_S$,
whilst, if $\alpha_i'>\alpha_3$, models (b), (g) imply lower string
scales than model (e).

The values of the string scale for models (a), (d) (represented in
the figure with dashed curves) are substantially higher; for these
latter cases in particular, assuming reasonable gauge coupling
relations $\alpha_i'\approx {\cal O}( \alpha_{2,3})$  we find that
$M_S\ge 10^{12}$GeV. Again, for $\alpha_3=\alpha_i'$, (the zero
value of the $x$-axis) we rederive the values of $M_S$ presented in
Table 2.

\section{Yukawa coupling evolution and mass relations}

In this section, we will examine whether a unification  of the $b
- \tau$ Yukawa couplings is possible in the above described low
string scale models\footnote{ For $b-\tau$ unification in a
different context see also~\cite{Parida:1996mz}.}. Our procedure
is the following:

Using the experimentally determined values for the third generation
fermion masses $m_b, m_{\tau}$ we run the 2-loop system of the
$SU(3)_C\times U(1)_Y$ renormalization group equations up to the
weak scale ($M_Z$) and reconcile there the results with the
experimentally known values for the weak mixing angle and the gauge
couplings. For the renormalization group running below $M_Z$ we
define the parameters
\begin{eqnarray}
\tila_e \;=\; \left( \frac{e}{4 \pi} \right)^2 ,\, \tila_3 \;=\;
\left( \frac{g_3}{4 \pi} \right)^2 ,\, t \;=\; 2 \ln\mu
\end{eqnarray}
where $e, g_3$ are the electromagnetic and strong couplings
respectively and $\mu$ is the renormalization scale. The relevant
RGEs are \cite{Arason:1991ic}%
\begin{eqnarray}
\frac{d\tila_e}{dt} &=& \frac{80}{9}\tila_e^2+
\frac{464}{27}\tila_e^3+\frac{176}{9}\tila_e^2\tila_3\nonumber
\\[2mm]
\frac{d\tila_3}{dt} &=& -\frac{23}{3}\tila_3^2 -
\frac{116}{3}\tila_3^3 + \frac{22}{9}\tila_3^2\tila_e -
\frac{9769}{54}\tila_3^4 \nonumber
\\[2mm]
\frac{dm_b}{dt} &=& m_b \left\{ -\frac{1}{3} \tila_e - 4\tila_3 +
\frac{397}{162}\tila_e^2 - \frac{1012}{18} \tila_3^2 - \frac{4}{9}
\tila_3 \tila_e - 474.8712 \tila_3^3 \right\}\nonumber
\\[2mm]
\frac{dm_\tau}{dt} &=& m_\tau \left\{ -3\tila_e +
\frac{373}{18}\tila_e^2 \right\}\nonumber
\end{eqnarray}
where $m_b, m_\tau$ are the running masses of the bottom quark and
the tau lepton respectively, while we use the notation $\tilde{a}_i
\equiv g^2_i/16\pi^2$ and $\tilde{a}_{t,b,\tau} \equiv
\lambda^2_{t,b,\tau}/16\pi^2$.

The required value for the running mass of $m_t$ at $M_Z$ is
computed as follows: we formally solve the 1-loop RGE system for
($\tilde{a}_3$, $\tilde{a}_2$, $\tilde{a}_Y$, $\tilde{a}_t$,
$\tilde{a}_b$, $\tilde{a}_\tau$) and afterwards we determine the
interpolating function for $\tilde{a}_3(\mu)$ and $m_t\left(\mu;
m_t(M_Z)\right)$ at any scale $\mu$ above $M_Z$, where $m_t(M_Z)$
indicates the dependence on an arbitrary initial condition. The
unknown value for $m_t(M_Z)$ is determined by solving numerically
the algebraic equation%
\ba%
\left[ m_t\left(\mu; m_t(M_Z)\right) - \frac{M_t}{1+ \frac{16}{3}
\tilde{a}_3(\mu)-2 \tilde{a}_t(\mu)} \right]_{\mu=M_t}=0
\ea%
We use these results as inputs for the relevant parameters and we
run the RGE system to the scale where the $\tilde{a}_b$ and
$\tilde{a}_{\tau}$ Yukawa couplings coincide. In our numerical
analysis we use for the gauge couplings the values presented in the
previous section, for the bottom quark mass $m_b$ the experimentally
determined range at the scale  $\mu = m_b$, i.e., ~$m_b(m_b) = 4.25
\pm 0.15$ GeV and finally the top pole mass is taken to be
$M_t=178.0 \pm 4.3$ GeV \cite{Eidelman}.

For the scales above $M_Z$ we consider the standard model spectrum
augmented by one more Higgs. The Higgs doubling is in accordance
with the situation that usually arises in the SM variants with brane
origin. Moreover, we assume that one Higgs $H_u$ only couples to the
top quark while the second Higgs $H_d$ couples only to the bottom.
Then, in analogy with supersymmetry we define the angle $\beta$
related to their vevs where $\tan\beta=\frac{v_u}{v_d}$. Thus, we
have the equations for the gauge couplings
\begin{eqnarray}
  \frac{d\tila_Y}{dt}\;=\;  7 \tila_Y^2 ,\;\;
  \frac{d\tila_2}{dt}\;=\; -3 \tila_2^2 ,\;\;
  \frac{d\tila_3}{dt}\;=\; -7 \tila_3^2 \nonumber
  \end{eqnarray}
  and for the Yukawas
  \begin{eqnarray}
  \frac{d\tila_t}{dt}&=&  \tila_t (-\frac{17}{12}\tila_Y
  - \frac{9}{4} \tila_2 - 8 \tila_3 + \frac{9}{2} \tila_t + \frac{1}{2} \tila_b)  \nonumber\\
  \frac{d\tila_b}{dt} &=& \tila_b (-\frac{5}{12} \tila_Y
  - \frac{9}{4} \tila_2 -8 \tila_3 + \frac{1}{2} \tila_t + \frac{9}{2} \tila_b + \tila_{\tau}) \nonumber\\
  \frac{d\tila_{\tau}}{dt} &=&  \tila_{\tau} (-\frac{15}{4} \tila_Y
  - \frac{9}{4} \tila_2 + 3 \tila_b + \frac{5}{2} \tila_{\tau} )   \nonumber\\
  \frac{dv_u}{dt} &=& \frac{v_u}{2} (\frac{3}{4} \tila_Y + \frac{9}{4} \tila_2
  - 3 \tila_t)           \nonumber   \\
  \frac{dv_d}{dt} &=& \frac{v_d}{2} (\frac{3}{4} \tila_Y + \frac{9}{4} \tila_2
  - 3 \tila_b - \tila_{\tau} )\nonumber
\end{eqnarray}
where  $ t = 2\ln\mu$.

Further, if we define $ v^2 = v_u^2 +v_d^2 $, with $v_u = v
\sin\beta $, $v_d = v \cos\beta$ and $v\sim 174~ {\rm GeV}$, the
$Z$-boson mass is given by $M_Z^2 = \frac{1}{2}(g_Y^2 + g_2^2)v^2$.
The elecromagnetic and the strong couplings are defined in the usual
way
\[ \tilde{\alpha}_{e} =  \tilde{\alpha}_Y \cos^2\theta_W =
\tilde{\alpha}_2 \sin^2\theta_W \] while the top and bottom quark
masses are related to the Higgs vevs by
\[ m_t = 4 \pi v_u \sqrt{\tilde{\alpha}_t} ~~~~~~~~ m_b = 4 \pi v_d \sqrt{\tilde{\alpha}_b} \]

We will examine the possibility of obtaining $b-\tau$ unification at
a low string scale $M_S$. We first concentrate in the models (a)-(g)
discussed in the previous section. We present our results in the
last column of Table \ref{ytab3a}.
\begin{table}[!ht]
\centering
\begin{tabular}{|c|c|c|c|c|}
\hline model & $k_Y$ & $\xi=\frac{\alpha_2}{\alpha_3}$ & ${M_S}/{GeV}$&$\frac{m_b}{m_{\tau}}(M_S)$\\
\hline
b,e,g & $2.969$ & 0.42 & $5.786 \times 10^3$ &1.25     \\
\hline
c     & $2.539$ & 0.58 & $1.986 \times 10^6$  &1.01    \\
\hline
d     & $1.554$ & 0.93 & $1.645 \times 10^{14}$  &0.73  \\
\hline
a     & $1.226$ & 1.01 & $6.710 \times 10^{17}$ &0.68   \\
\hline
\end{tabular}
\caption{\label{ytab3a} The  String scale and the ratio
$\xi=\frac{\alpha_2}{a_3}$ for various orientations of $U(1)$ branes
presented  in Table \ref{ytab2}. The last column shows the $b-\tau$
ratio at $M_S$. Exact $b-\tau$-unification is obtained in model (c)
for $M_S\sim 10^3TeV$.}
\end{table}

We notice that $b-\tau$ unification is obtained in model (c), for
$M_S\approx 2\times 10^6$ GeV. Models (b,e,g) with unification scale
$M_S\approx 5.8 \times 10^3$ GeV predict a small (25\%) deviation
from exact $b-\tau$ unification. We observe that in these cases the
strong-weak gauge coupling ratio is ${a_3}\approx 2\,{a_2}$. As
noted previously, the relation $\alpha_3=2\,\alpha_2$ holds
naturally if we embed the model in a $U(3)\times U(2)^2\times
U(1)^2$ symmetry.

\begin{figure}[!ht]
\centering
\includegraphics[width=0.9\textwidth]{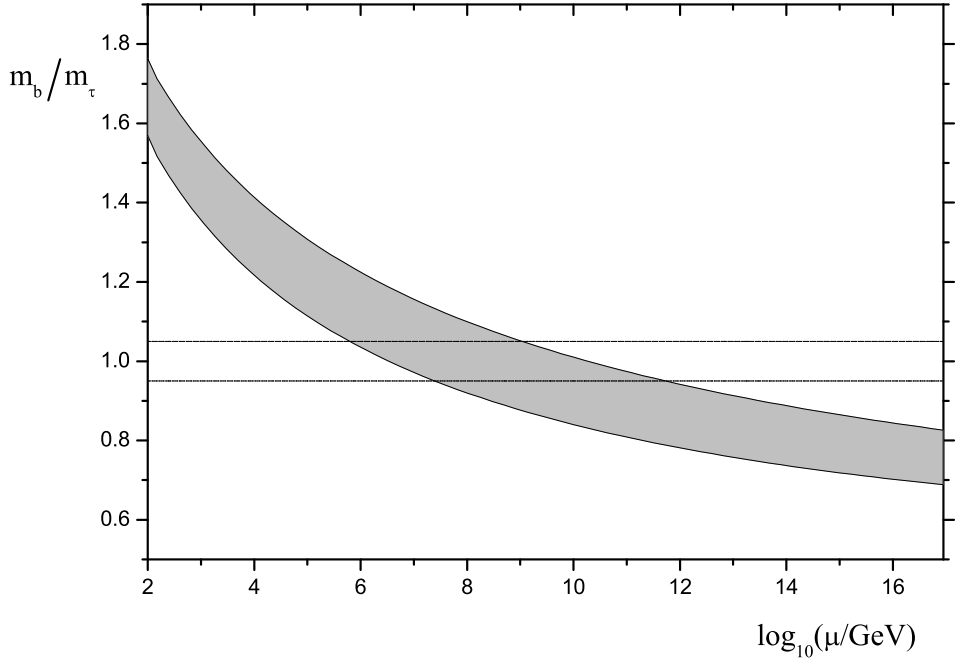}
\caption{\label{mbmtau}The ratio $\frac{m_b}{m_{\tau}}$ as a
function of the energy $\mu$ in the 2-Higgs Standard Model. The
shaded region corresponds to $a_3$ uncertainties. The two horizontal lines
indicate the interval  $\pm 5\%$ around the unity.}
\end{figure}

In figure \ref{mbmtau} the ratio $\frac{m_b}{m_{\tau}}$ is plotted
as a function of the energy scale for the case of the two-Higgs
Standard Model\footnote{For recent work on a 2 Higgs model
see~\cite{Kane:2005va} and references therein.}.
 All previous uncertainties are incorporated and
the result is the shaded region shown in the figure. The horizontal
shaded band is defined between the values
$\frac{m_b}{m_{\tau}}=[0.95-1.05]$ taking into account deviations of
the ratio $\frac{m_b}{m_{\tau}}$ from unity due to possible
threshold as well as mixing effects in the full $3\times 3$ quark
and lepton flavor mass matrices.
%Figure 4
As can be seen, exact $m_b=m_{\tau}$ equality is found  around the
scale $M_S\approx 2\times 10^6$ GeV. Taking into consideration
$m_b/m_{\tau}$-uncertainties expressed through the shaded band, the
$M_S$ energy range is extended up to $\sim 10^{12}$ GeV.

\section{Conclusions}

In this letter, we performed a systematic study of the Standard Model
embedding in brane configurations with $U(3)\times U(2)\times U(1)^N$
gauge symmetry and we  examined a number of interesting phenomenological
issues. Seeking for models with economical Higgs sector, we identified two
brane configurations with two or three ($N=2,3$) $U(1)$-branes which can
accommodate the Standard Model where only one Higgs doublet couples to the
up quarks, and a second one couples to the down quarks and leptons. We
analysed the possible hypercharge embeddings and found seven possible solutions
leading to six models (with acceptable string scale $M_S$), implying the correct
charge assignments for all standard model particles.

We further examined the gauge coupling evolution in these models for
both, parallel, as well as intersecting branes and determined the
lowest string scale allowed for all possible alignments of the
$U(1)$ branes with respect to the $U(3)$ and $U(2)$ non-abelian
factors of the gauge symmetry. In the parallel brane scenario, we
have identified three models which allow a string scale $M_S$ as low
as a few TeV, one model  with string scale of the order $10^6$ GeV
and two models  with high unification scales. Similar results were
obtained for the general case of intersecting branes.

We further analysed the consequences of the third generation fermion mass relations
and in particular $b-\tau$ equality at the string scale on the above models.
In the parallel brane scenario, we found that exact $b-\tau$ Yukawa unification
is obtained only in the model with $M_S\approx 10^3$TeV, while in the TeV  string
scale models  the $m_b/m_{\tau}$ ratio deviates from unity by $25\%$.
Allowing the $U(1)$ gauge couplings to take arbitrary (perturbative) values, we found that $b-\tau$
Yukawa unification is possible for a wide string scale range form $10^6$ up to $10^{12}$ GeV.
\\[2cm]%
{\bf Acknowledgements}. {\it This research was funded by the program
`PYTHAGORAS' (no.\ 1705 project 23) of the Operational
 Program for Education and Initial Vocational Training of the Hellenic
 Ministry of Education under the 3rd Community Support Framework and
 the European Social Fund. GKL and JR would like to acknowledge kind hospitality
 of Ecole Normale Superiere in Paris, where part of this work has been done.}

\end{document}